\documentstyle[epsf,fleqn,aps]{revtex}
\begin{document}
\title{Equilibrium of anchored interfaces with quenched disordered growth}
\author{M. D. Grynberg}
\address{Departamento de F\'{\i}sica, Universidad Nacional de La Plata,
(1900) La Plata, Argentina}
\maketitle
\maketitle

\begin{abstract}
The roughening behavior of a one-dimensional interface fluctuating 
under quenched disorder growth is examined while keeping an anchored 
boundary. The latter introduces detailed balance conditions which allows 
for a thorough analysis of equilibrium aspects at both macroscopic and 
microscopic scales. It is found that the interface roughens linearly with
the substrate size only in the vicinity of special disorder realizations.
Otherwise, it remains stiff and tilted.

\vspace{10 pt}


\vspace{-12 pt}
\end{abstract}

\vskip 0.2cm
\vskip2pc

Studies of inhomogeneous interface growth have thrived in a variety of 
physical contexts \cite{Krug} characterizing phenomena as diverse as fluid
imbibition in porous media \cite{Amaral} and crystalline surface growth 
on disordered substrates \cite{Shapir}. It is known that in those 
situations a small amount of disorder can severely modify the interface 
motion and ultimately alter its roughening behavior \cite{Krug}. This is 
the case of time independent but spatially random growth rates arising, 
for instance, from a quenched array of columnar defects pinning the flux 
lines, here playing the role of interfaces, in dirty high temperature 
superconductors \cite{Balents}. Another mechanism whereby the interface 
character results deeply affected at large times is realized by anchoring 
conditions which suppress fluctuations at the interface boundaries.
Such confined geometries actually occur in the unbinding of polymers 
from a wall \cite{Forgacs}, and also emerge as domain walls of (2+1) 
dimensional cellular automaton models \cite{Derrida}, as well as in 
stationary nonequilibrium systems in $d = 1\,$ \cite{Krug2}. In this 
work we focus on the combined effect that quenched disorder and anchoring
conditions brings about in the steady state (SS) properties of $1d$ 
interfaces, at both macroscopic and microscopic scales.

The problem is most conveniently treated in the discrete formulation of 
growth processes. As usual \cite{Plischke}, here we represent the latter 
in terms of restricted solid on solid (RSOS) configurations of heights 
$(h_0),\,h_1,\,...\,, h_L$ growing stochastically on a substrate of size 
$L$. Throughout the evolution, fluctuations are suppressed entirely at 
$h_0$, whereas all other $h_j$ can increase (decrease) in two height 
units with substrate dependent rates $\epsilon_j$ ($\epsilon'_j$) 
\cite{other}. Due to the RSOS constraints $\vert h_{j+1} - h_j \vert = 1$,
these variations can only occur at local extrema of the interface, 
as illustrated in Fig. 1. Despite its simplicity, it will turn out that 
this model encompasses a disorder driven transition along with unusual 
roughening exponents.

Following most studies \cite{Krug}, we concentrate on the root mean square
dispersion of heights, commonly associated to the width $W\,$ of the 
system. Under quenched growth this further requires the averaging of $W$ 
over all disorder realizations (denoted by the brackets below), namely
\begin{equation}
\label{width}
\langle\, W (L,t) \,\rangle =  \Big\langle\, \Big\{
\frac{1}{L} \,\sum_j\, \overline{\left[ \, h_j(t) -  H(t) \,\right]^2}
\Big\}^{1/2} \Big\rangle\,,
\end{equation}
where $H (t)$ is the mean interface height, and the overbar indicates an 
ensemble average over all possible evolution histories up to time $t$. 
On general grounds \cite{Krug,Family}, it can be argued that $\langle\, 
W (L,t) \,\rangle\,$ scales as $L^\zeta f (t/L^z)\,$ with a universal 
scaling function behaving as $f(x) \sim x^{\zeta/z}\,$ for $x \ll 1\,$, 
whereas for $x \gg 1\,$ it remains constant. Consequently, for $t \gg 
L^z\,$ the width saturates as $L^\zeta\,$ while growing as $t^{\zeta/z}$ 
at much earlier stages. In what follows we content ourselves with studying
just the former situation i.e. the stationary regime controlled by the 
roughening exponent $\zeta\,$. In contrast to the dynamic exponent $z$,\,
$\zeta$ lends itself more readily for a thorough analysis under growth 
disorder, so hereafter we shall work over Eq.\,(\ref{width}) directly in 
the limit $t \to \infty\,$.

{\it a. Particle representation} -- 
To this aim, let us first consider the conditions that the SS imposes on
the above processes. For ease of analysis we retain the interface slopes 
$s_j = h_j - h_{j-1} = \pm 1$ rather than its heights, or alternatively
the set of occupation numbers $\{n\}\equiv \{(1+s)/2\}$ conforming a 
height configuration. This enables to exploit the well known mapping 
\cite{Krug,Plischke} from a $1d$ RSOS interface to a driven particle 
system, i.e. $h_j = \sum_{i\le j} (2n_i - 1)$, here specially adapted to 
account for both anchored and free edges as well as for quenched disorder.
In particular, height variations at the free boundary translate here into
injection and ejection of particles, as is easily visualized in Fig. 1. 
In this latter representation one can readily build up a SS probability 
measure $P (\{n\})\,$ satisfying detailed balance, regardless the specific
realization of disorder considered. In equilibrium, the bulk and right 
boundary processes will demand respectively (see Fig. 1)
\begin{eqnarray}
\label{bulk}
\epsilon_i' \, P (\, n_1,\, ... \, \underbrace{1,0}_{i,i+1},
\, ... \,n_L ) &=& \epsilon_i \, P (\, n_1,\, ... \, 
\underbrace{0,1}_{i,i+1},\, ... \,n_L )\,,\\
\label{boundary} 
\epsilon_L' \, P (\, n_1,\, ... \, n_{L-1}, 1\,) &=& \epsilon_L \, 
P (\, n_1,\, ... \, n_{L-1},0\,)\,,
\end{eqnarray}
and therefore the SS distribution must be of the form $e^{- \sum_j 
\! V_j n_j}$. Specifically, (\ref{bulk}) entails a hard-core particle 
potential $V_j = V_1 + \sum_{i < j}\, \ln \left(\epsilon_i/\epsilon'_i
\right)$, whereas (\ref{boundary}) fixes the value of the additive
constant. Equivalently, by introducing the total particle number 
${\cal N} = \sum_j n_j\,$, we thus obtain a sample dependent grand 
canonical form
\begin{equation}
\label{grand}
P (\{n\}) = \frac{\mu^{\cal N}}{Z}\,
\, \exp \,\Big( - \sum_j\, V_j \,n_j\,\Big)\,,
\end{equation}
($V_1 \equiv 0\,$), with a particle fugacity $\mu = \prod_j \,
\epsilon_j/\epsilon'_j\,$ and a normalizing partition function $Z = 
\prod_j ( 1 + \mu\, e^{-V_j} )\,$. 

This provides the basic elements to compute the original average 
(\ref{width}). In this regard and for posterior use, it can be easily 
checked that the particle densities $\overline n_i\,$ involved in the 
calculation of height profiles are
\begin{equation}
\label{dens}
\overline n_i = \Big(\,1 + \prod_{j \ge i} \,\epsilon'_j/\epsilon_j\,
\Big)^{-1}\,,
\end{equation}
while the pair correlations needed for the 
analysis of height-height correlators appearing in $W\,$, result totally 
decoupled, that is, $\overline {n_i\,n_j} = \overline n_i \, \overline 
n_j\;\, \forall\, i \ne j\,$ and for all disorder realizations. However as
we shall see below, there is a particular situation, namely $\mu \to 1\,$,
for which height variables become strongly correlated.

{\it b. Height distribution} -- 
Before embarking in the evaluation of the double average of 
Eq.\,(\ref{width}), we should determine firstly whether the growth of 
a given sample can actually produce a rough interface within the 
equilibrium regime (\ref{grand}). It may well happen that while the width 
increases with the typical substrate size, the whole system becomes 
actually smooth, as is the case of the linear profile exhibited in Fig. 1.

In analyzing this issue we focus attention on a more microscopic level of
description such as the {\it single} height probability density $P(h_N)\,$
at a given location $N$. In turn, this requires the evaluation of an 
$M$-particle partition function $Z_M \equiv \omega_{_M}/M!\,$ 
constructed as  
\begin{equation}
\label{partition}
\omega_{_M}  = \mu^M \!\!\!\!\!\!\sum_{j_{_1} ...\, j_{_M} \le _N}
\!\!\!\!\!\!\!\!' \,e^{\,-V_{j_1}} ...\, e^{\,- V_{j_{_M}}}\,,
\end{equation}
where $\sum\!'$ restricts the sums to $\{ j_{_1} \ne \,...\, 
\ne j_{_M} \} \in [1,N]\,$ so as to keep a constant number of particles 
within that interval. Clearly, this $M$-body quantity ($M \le N\,$) is 
associated to the wanted probability at $h_N = 2 M - N\,$, since by 
construction $P(2 M - N) = Z_M /[\,\prod_{j=1}^N (1 + \mu e^{-V_j})\,]\,$.
To account for the hard-core interactions we build up a recursive 
relation in the particle number $M$. Introducing the auxiliary functions 
$g (k) = \mu^k \sum_{j=1}^N e^{- k\,V_j}\,$, this can be achieved via the
following identity
\begin{equation}
\omega_{_M}  = g (1) \, \omega_{_{M-1}} \,-\,(M-1) \,S_{_{M-2}} (2)\,,
\end{equation}
where the latter  factor is defined generically as $S_{_{M-k}} (k) = 
\mu^M \,\sum\!'_j \: e^{\,- k\,V_j} \sum\!'_{j_{_1} ...\,j_{_{M-k} }}
e^{\,- V_{j_{_1}}} ... \, e^{\,- V_{j_{_{M-k}}}}\,$. In particular, 
$S_{_{M-2}} (2)$ involves constrained sums having $\{j \ne j_1 \,...\,
\ne j_{M-2} \} \in [1,N]\,$ and subtracts the unwanted terms resulting 
from $(M-1)$ double occupation configurations included in $g (1)\,
\omega_{_{M-1}} \,$. Using a similar criteria, we may also infer that 
$S_{_{M-2}} (2) = g (2)\,\omega_{_{M-2}} - (M-2)\,S_{_{M-3}} (3)\,$,
where the last term now cancels triple occupation contributions to 
$S_{_{M-2}} (2)\,$. Thereafter, iterating this reasoning down to $M=1$, 
it can be straightforwardly verified that in terms of the $j$-particle 
partition functions $Z_j = \omega_{_j}/j!$ we are left with
\begin{equation}
\label{recursion}
Z_M = \frac{(-1)^{M+1}}{M}\,\Big[ g (M) + 
\sum_{j=1}^{M-1}  (-1)^j \, g (M-j) \,Z_j \Big], 
\end{equation}
where $Z_1 = g(1)\,$. Although the analytic solution of such recurrence 
is not reachable by standard means \cite{Wilf}, nevertheless it allows 
for a simple numerical evaluation of the height distribution, particularly
at the tails where statistics becomes more demanding. Despite the noisy  
particle potential $V$ of individual samples it turns out that for 
$M \gg 1\,$ the distribution itself collapses towards a universal 
-Gaussian- form. This is evidenced by the semi-log inset of Fig. 1, for 
instance, in the case of a binary concentration $p$  of growth rates with 
probability  $p\, \delta ({\epsilon' \over \epsilon} - a) + (1-p) \,
\delta ({\epsilon' \over \epsilon} - b)\,$. Also, recursion 
(\ref{recursion}) reveals that in most cases the interface is 
characterized by a {\it tilted} profile $\overline h_N \sim \pm N\,$ 
whose local height fluctuations (HF) set out to be quite {\it small}, 
even far away from the anchored boundary $h_0\,$, i.e. the interface is 
{\it stiff}, in agreement with the linear snapshot displayed in Fig. 1.
On approaching certain disorder conditions however, Gaussian tails widen 
significantly (slope diminishing shown in the inset), so HF take over
and the stiff regime no longer holds. This is illustrated by the rough 
profile also exhibited in Fig. 1.

{\it c. Roughening criterion} -- 
It is therefore important to focus on these
special conditions so as to ensure a meaningful result in the calculation
of (\ref{width}). Let us then consider the fluctuations of the total 
number of particles $\langle\, \sigma^2\rangle \equiv \langle \overline 
{{\cal N}^2} - \overline {\cal N}^2 \rangle \,$, averaged over a generic 
growth disorder. For equilibrium regimes, this is a representative 
quantity to look at, as in terms of the occupation densities discussed 
above it is immediate to check that it coincides with the height variance 
at the free edge of the interface, i.e. $4\,\langle\, \sigma^2\rangle = 
\langle\, \overline {h_L^2} - \overline h_L^2 \big\rangle\,$ whereas by 
construction it constitutes an upper bound for all HF. In turn, due 
to the RSOS constraint this variance also provides a measure of 
fluctuations deep inside the bulk. To facilitate its analysis and by 
virtue of the products involved in Eq.\,(\ref{dens}), it is convenient 
to introduce the rate disorder variables $u_i = \ln \big (\epsilon'_i/
\epsilon_i \big)\,$, in terms of which $\langle\,\sigma^2 \rangle\,$ 
adopts the form
\begin{equation}
\label{fluc}
\langle\, \sigma^2 \rangle = \frac{1}{4}\,\sum_{j=1}^L\,\left\langle\,
\left[\,\cosh \Big(\,\frac{1}{2}\sum_{i=j}^L u_i \Big)\,\right]^{\!-2}
\right\rangle\,.
\end{equation}
Assuming a site independent distribution of growth ratios, say with mean 
$\langle u \rangle$ and variance $s^2\,$, we can thus exploit the 
central limit theorem \cite{Feller} for these new variables and 
carry out the disorder average. Hence, it can be readily shown that 
\begin{equation}
\label{central}
\langle \, \sigma^2 \rangle \simeq 
\sum_{j=1}^k \alpha_j\, + \,\frac{1}{4 s }\,\sum_{j=k+1}^{L}\; 
\int \limits_{-\infty}^{+\infty}\;
\frac{\exp\,\left[ -\frac{(u - j \langle u \rangle)^2}
{2 \, j \, s^2} 
\right]}{\sqrt{2 \pi j\, } \,\cosh^2 (\frac{u}{2})}\; d u\,,
\end{equation}
where $ \alpha_j = \langle\,\overline n_{_{L+1-j}}\, (1 - \overline 
n_{_{L+1-j}})\,\rangle\,$, and  $k\,$ is eventually a large but 
{\it finite} integer ($ k \ll L$). In particular, for a Gaussian 
$u\,$-distribution $\langle \, \sigma^2 \rangle\,$ just reduces to the 
integrals sum ($k=0\,)$. The point to emphasize here is that whatever
distribution is considered in (\ref{fluc}), the interface fluctuations 
are intrinsically dominated by the above series of integrals. Thus, apart 
from a bounded quantity $0 \le \sum_{j=1}^k \alpha_j \le k/4$, the effect 
of any distribution on $\langle\,\sigma^2 \rangle\,$ will parallel the 
one caused by the Gaussian disorder. On the other hand, as long as 
$s \ne 0\,$ the asymptotic analysis of (\ref{central}) shows that the 
value of $\langle u \rangle$ is crucial, since for large $L$ 
\begin{equation}
\label{condition}
\langle\, \sigma^2 \rangle \propto 
\cases{ \sqrt L\,, \;\;\;\, {\rm if} \;\; \langle\, \ln 
\frac{\epsilon'}{\epsilon} \rangle = 0\,, \cr
{\rm bounded, \;\; otherwise}\,, }
\end{equation}
which finally provides the disorder conditions able to produce a rough 
interface. As was mentioned earlier, this occurs for fugacities such that 
$\mu \to 1^{\pm}\,$ in turn favoring strong fluctuations in the occupation
numbers of Eq.\,(\ref{grand}). In that case, $\vert \overline s_i \vert 
< 1\,$ and height variables become tightly correlated along the substrate 
because $\overline {h_i\,h_j} - \overline h_i \, \overline h_j = (\,i - 
\sum_{k=1}^i \overline s_k^2\,) > 0,\, \;\, \forall\, i < j\,$. Otherwise,
the slopes $\overline s_i \to -1, 1\,$ (i.e. $\bar {\cal N}/L \to 0,1\,$),
so the interface becomes tilted and hardly flexible since as stressed 
before $\langle\,\overline {h_i^2} - \overline h_i^2 \big\rangle \le 4 \,
\langle\, \sigma^2 \rangle\,$ $\forall\, i\,$.

To illustrate the effectiveness of the above criterion for non-Gaussian 
cases, we address to the binary distribution referred to above. Using the 
binomial distribution to weight all possible forms in which the growth 
ratios $a,\,b\,$ may show up on a segment of length $l\,$, we find
\begin{equation}
\label{flucb}
\langle\, \sigma^2 \rangle = \sum_{l=1}^L\,\sum_{k=0}^l\,
{l \choose k}\, \frac{p^k\,(1-p)^{l-k} }{4 \,\cosh^2 \varphi_{l,k} }\,,
\end{equation}
where $2\,\varphi_{l,k} \equiv k \ln a + (l-k) \ln b\,$. Fig. 2 shows the 
typical behavior of this quantity for $a < 1 < b\,$. In agreement with 
condition (\ref{condition}), there exists a critical concentration 
$p_c = [\,1 - \ln a/\ln b\,]^{-1}\,$, i.e. $\langle u \rangle = 0\,$, 
for which particle fluctuations increase as $\sqrt  L\,$ (see inset of 
Fig. 3), while  in the thermodynamic limit diverge as $\vert p - p_c 
\vert^{-1}\,$ as indicated by the inset of Fig. 2. In nearing $p_c$ from 
below (above), this disorder-driven transition brings the interface 
from a tilted state with slope $-1\,$ ($+1\,$) to a rough phase in a 
continuous manner. In this respect, size effects become quite noticeable 
when passing through $p_c\,$. Clearly, for $b < 1 < a\,$ the rapprochement
direction is inverted, otherwise the system remains tilted. 

It is of interest to note here that criterion (\ref{condition}) implies 
that locally, uniform interfaces ($s = 0\,$) can fluctuate stronger than 
disordered ones. In fact, for $\epsilon'_i = \epsilon_i\,$ all interface 
configurations are equally likely so the single height probability density
derived from Eq.\,(\ref{partition}) reduces to $P(h) \sim \frac{1}{\sqrt
 L} \exp (-\frac{h^2}{2L})\,$. So, in the pure substrate $\sigma^2 
\propto L\,$, rather than diverging as in (\ref{condition}). What remains
to be seen is whether different scaling behaviors can also emerge on a 
more macroscopic description, such as the width of Eq.\,(\ref{width}) and
towards which we finally turn. 

{\it d. Width behavior} -- 
Let us then consider this global quantity in the slope representation 
discussed throughout. After some algebraic steps, it can be readily 
demonstrated that the width of a given sample is expressed as
\begin{equation}
\label{width2}
W^2 (L) = \frac{\left( L^2 - 1 \right)}{6\,L}\,+\,
\frac{2}{L^2}\,\sum_{i < j}\,(L+1-j)\,(i-1)\; \overline {\,s_i s_j}\,.
\end{equation}
In particular, for equilibrium regimes there is a key simplification in 
the average of $W$, as $\overline {\,s_i s_j} = \overline s_i \overline
s_j\,$ over all disorder realizations. In passing, it is worth recalling 
that in a uniform unanchored system $W \sim \sqrt {L/12}\,$ 
\cite{Plischke}, thus for $\epsilon_i = \epsilon'_i\,$ where $\overline 
s_i \equiv 0\,$, the anchored boundary just modifies a mere amplitude.

For binary rates, a closed expression of $\langle\, \overline s_i 
\overline s_j\rangle\,$ can be easily found upon recurring once more to 
the binomial distribution. According to the occupations (\ref{dens}), 
the contributions of $\overline s_i \overline s_j\,$ to its own quenched 
average are simply obtained by weighting the joint number of possibilities
in which the growth ratios $a, \, b\,$ appear along the intervals 
$[i,j-1]\,$ and $[j, L]\,$. Specifically, this yields
\begin{eqnarray}
\nonumber
\left\langle\, \bar s_i \bar s_j\, \right\rangle &=& 
\sum_{k=0}^{L+1-j}\,\sum_{k'=0}^{j-i}\,
{L+1-j \choose k}\, {j-i \choose k'} \,\Big(\frac{p}
{1-p}\Big)^{k+k'} \\
&\times&\,(1-p)^{L+1-i}\,\tanh \theta_{j,k}\;\tanh \theta_{i,k+k'}\,,
\end{eqnarray}
where  $k$ ($k'\,$) is the number of $a$-occurrences within $[j, L]\,$ 
(\,$[i,j-1]\,$), while the $\theta$-arguments are generically defined as 
$2\,\theta_{j,k} \equiv k \ln a - (j+k-L-1) \ln b\,$. 
The size dependence of the resulting quenched width is plotted in Fig. 3
(solid lines), using the critical concentrations $p_c$ required by 
criterion (\ref{condition}). Clearly, the results support a power law 
growth of $\langle\, W \rangle\,$ consistent with a rather unusual 
roughening exponent $\zeta \simeq 1\,$ extended over more than three 
decades.  Thus, we see that in sharp contrast to what occurs at the
microscopic scale where HF of disordered interfaces are weaker 
(\,$\langle\,\sigma^2 \rangle \propto \sqrt L\,$) than uniform ones 
($\sigma^2 \propto L\,$), macroscopically the former can grow rougher
than the latter.

Notice however that on approaching weakly disordered regimes size effects
become quite pronounced. Even for large substrate lengths $\langle \,W 
\rangle$ grows with the $\zeta = 1/2$ exponent characteristic of 
homogeneous systems  \cite{Plischke,KPZ}\, though asymptotically the 
linear scaling is recovered. This suggests a rather abrupt change 
separating the uniform case from the more general disordered situation, 
namely: at the static level of the $\zeta\,$-exponents there is a 
{\it discontinuous} scaling regime at $p_c = 0^+,\,1^-\,$ where $\epsilon
\simeq \epsilon'\,$ for most sites. In fact, the numerical estimation of 
$\langle \,W\rangle\,$ via Eqs.\,(\ref{dens}) and (\ref{width2}) over a 
Gaussian growth distribution satisfying  $\big\langle \ln \big 
(\epsilon'/\epsilon \big)\,\big\rangle = 0\,$, as required by our 
criterion, yields the same value of $\zeta\,$ while also implying a 
discontinuous behavior near the homogeneous case, when $s \to 0^+\,$. 
This is indicated by the dashed lines of Fig. 3 after averaging over 
$10^4$ samples. Whether such discontinuous feature corresponds to a 
general aspect of weakly disordered roughening processes in $d = 1\,$ 
seems difficult to elucidate rigorously \cite{yo}.

To summarize, we have constructed a recursive relation 
[\,Eq.\,(\ref{recursion})\,] which allowed us to examine disordered 
interfaces at microscopic scales. For most situations this yielded a stiff
picture given the small HF obtained, though in the vicinity of special 
disorder realizations HF take over and wipe out the interface stiffness. 
This led us to propose a roughening criterion by looking at the 
fluctuations of the total particle number along with their divergence 
conditions [\,Eq.\,(\ref{condition})\,]. Within such regimes we analyzed
the scaling behavior of quenched widths using both Gaussian and non 
Gaussian disorders. The corresponding results support a linear growth of 
$\langle \,W \rangle\,$ with the substrate size, i.e. $\zeta \simeq 1\,$ 
(Fig. 3), which possibly reflects a tendency of anchored interfaces to 
crumple on large scales \cite{Krug,yo}.

Finally, though the non-equilibrium dynamics is out of the scope of this 
work, one could expect that criterion (\ref{condition}) will introduce 
somehow a crossover between ergodic and non-ergodic evolutions. It is then
natural to ask whether the fundamental scaling between length and time 
-\,embodied in the $z$ exponent of $W\,$- would be affected by our 
roughening condition. So far, this remains an open issue.

The author acknowledges support of CONICET, Argentina.

\begin{twocolumn}


\begin{figure}
\hbox{%
\hspace{-0.8cm}
\epsfxsize=4.1in
\epsffile{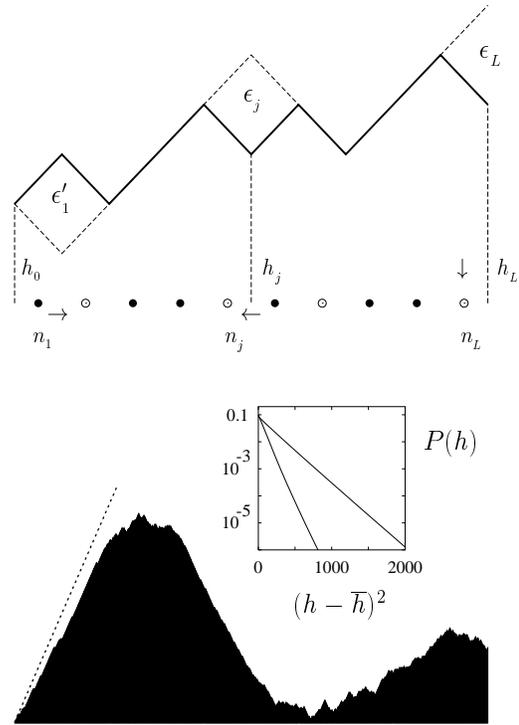}}
\vspace{-2cm}
\caption{RSOS interface anchored at $h_0$, thought of as an asymmetric
exclusion process in which height slopes $s_j = h_j - h_{j-1}$ 
are associated to particle occupations $n_j = \frac{1}{2}(1+s_j)$. 
At the free edge  $h_L$ variates in 2 ($-2$) with rate $\epsilon_L$ 
($\epsilon'_L)$ by particle deposition (evaporation), whereas for 
$1 \le j < L$ height changes correspond to left (right) particle hoppings
with rates $\epsilon_j$ ($\epsilon'_j$). Lower panel: typical snapshot 
for $L=10^3$ after $10^5$ steps per height using a binary disorder with 
$\epsilon'\!/\epsilon = 0.8,\, 1.2$ under condition (\ref{condition}),
i.e. for $p=p_c$ in the text.  Otherwise, configurations become tilted 
(dotted line). Inset: Gaussian distribution of heights for $j=100$,
at and slightly below $p_c$ (upper and lower lines).}
\end{figure}

\newpage
$\phantom{}$
\vskip -4cm
\begin{figure}
\hbox{%
\hspace{-1cm}
\epsfxsize=4.2in
\epsffile{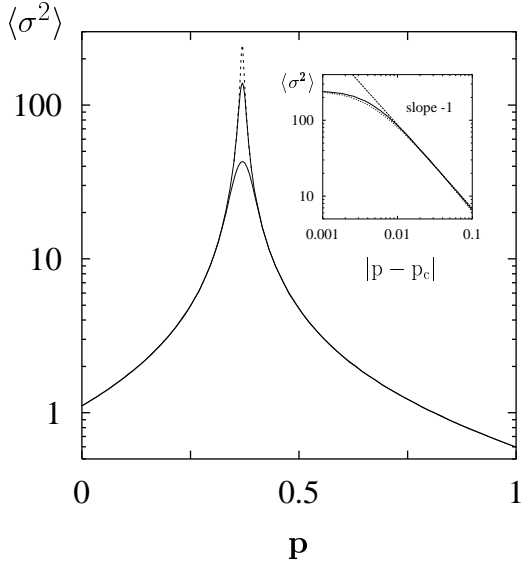}}
\vspace{-3.5cm}
\caption{Particle fluctuations arising from a binary growth distribution. 
Data for $b = 1.5\,$ while varying the concentration of $a = 0.5\,$. 
Curves from top to bottom denote in turn results for $L = 10^5, 10^4$ and 
$10^3$. The inset exhibits the same algebraic divergence (dashed slope)
either nearing $p_c = [\,1 - \ln a/\ln b\,]^{-1}\,$ from below or above 
(dotted and solid lines closely following each other; $L = 10^5$).}
\end{figure}

$\phantom{}$
\vskip -4cm
\begin{figure}
\hbox{%
\epsfxsize=4.2in
\hspace{-1.3cm}
\epsffile{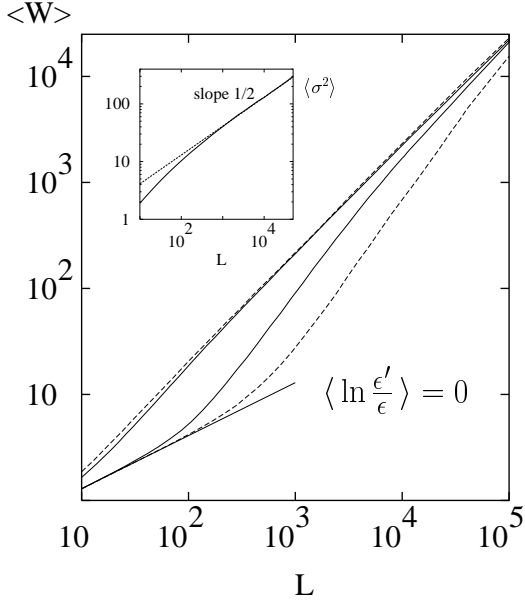}}
\vspace{-3cm}
\caption{Quenched interface width scaling linearly with the substrate 
size. Averages were taken over Gaussian and binary distributions (dashed 
and solid lines respectively), using condition (\ref{condition}). Curves 
in descending order stand in turn for $s = 1\,$, $p_c = 0.4\,$, 
$s = 0.01\,$ and $p_c = 0.004\,$ ($a = 0.5\,$); the lowermost slope is 
$1/2$. The inset displays fluctuations (\ref{flucb}) scaling as 
$L^{1/2}$ (dashed slope) for $p_c = 0.4$.}
\end{figure}

\end{twocolumn}

\end{document}